\pdfoutput=1
\documentclass[sigconf,screen,nonacm]{acmart}

\usepackage[subtle]{savetrees}
\usepackage{algorithm}
\usepackage{algpseudocode}
\usepackage{graphicx}
\usepackage{textcomp}
\usepackage{xcolor}
\usepackage{cleveref}
\usepackage{xspace}
\usepackage{subcaption}
\usepackage{booktabs}
\usepackage{multirow}
\usepackage{comment}
\usepackage{enumitem}
\usepackage{tikz}
\usepackage[export]{adjustbox}
\usepackage{float}
\usepackage{makecell}
\usepackage{tcolorbox}
\usepackage{changepage}
\usepackage{soul}
\sethlcolor{lightgray}
\begin{document}
\newcommand{\mitlogo}{\hspace{1pt}\includegraphics[height=0.8em]{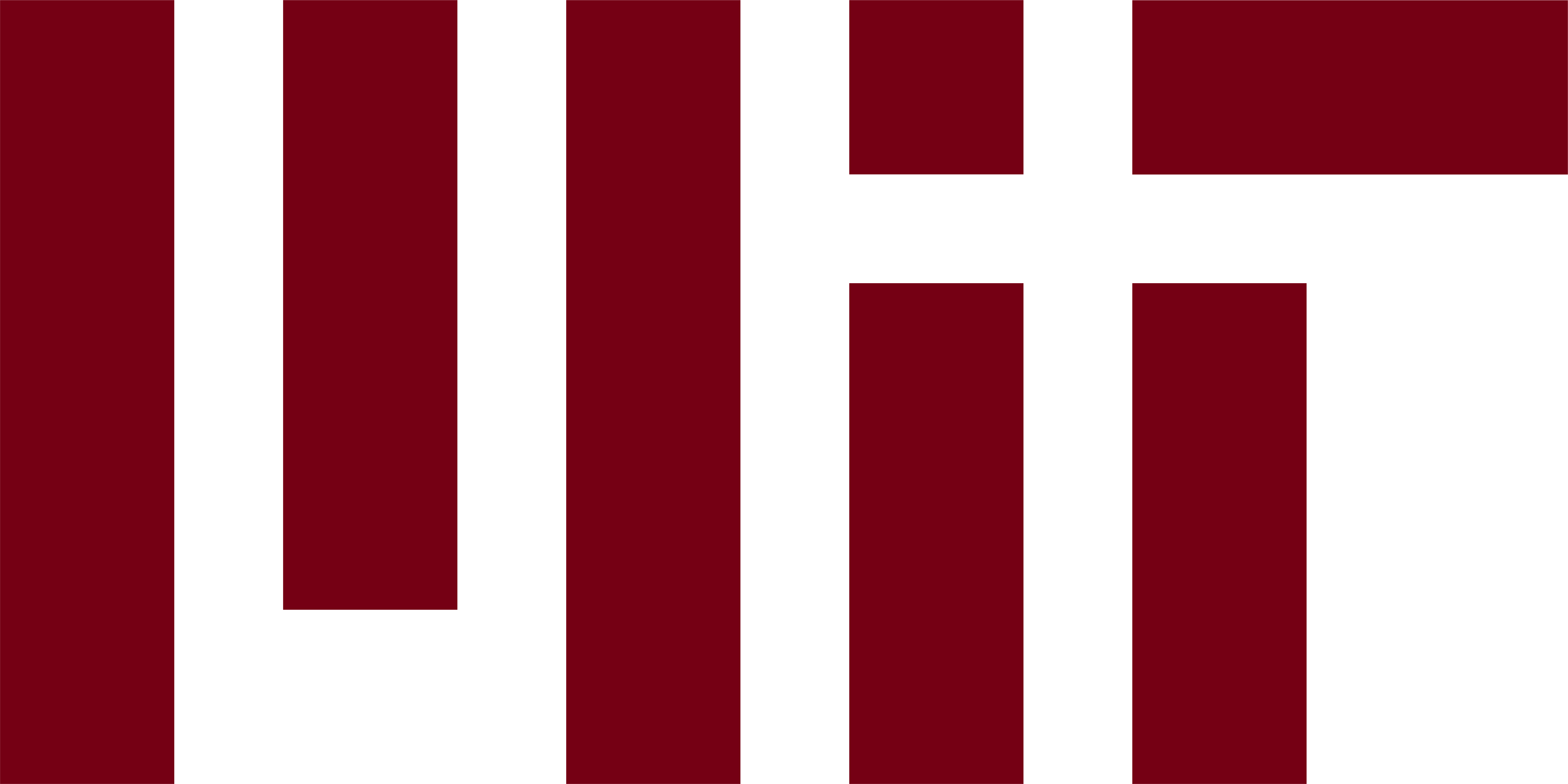}}
\newcommand{\googlelogo}{\hspace{0.4pt}\includegraphics[height=0.8em]{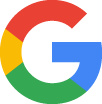}}
\newcommand{\nospacegooglelogo}{\includegraphics[height=0.8em]{logos/google.png}}
\newcommand{\uwlogo}{\hspace{0.2pt}\includegraphics[height=0.8em]{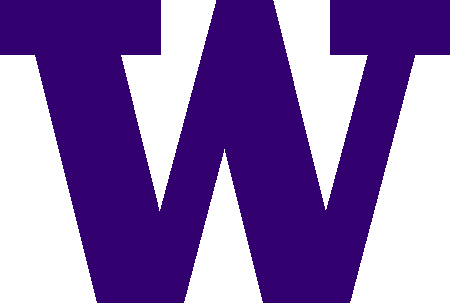}}
\newcommand{\rutgerslogo}{\hspace{1pt}\includegraphics[height=0.8em]{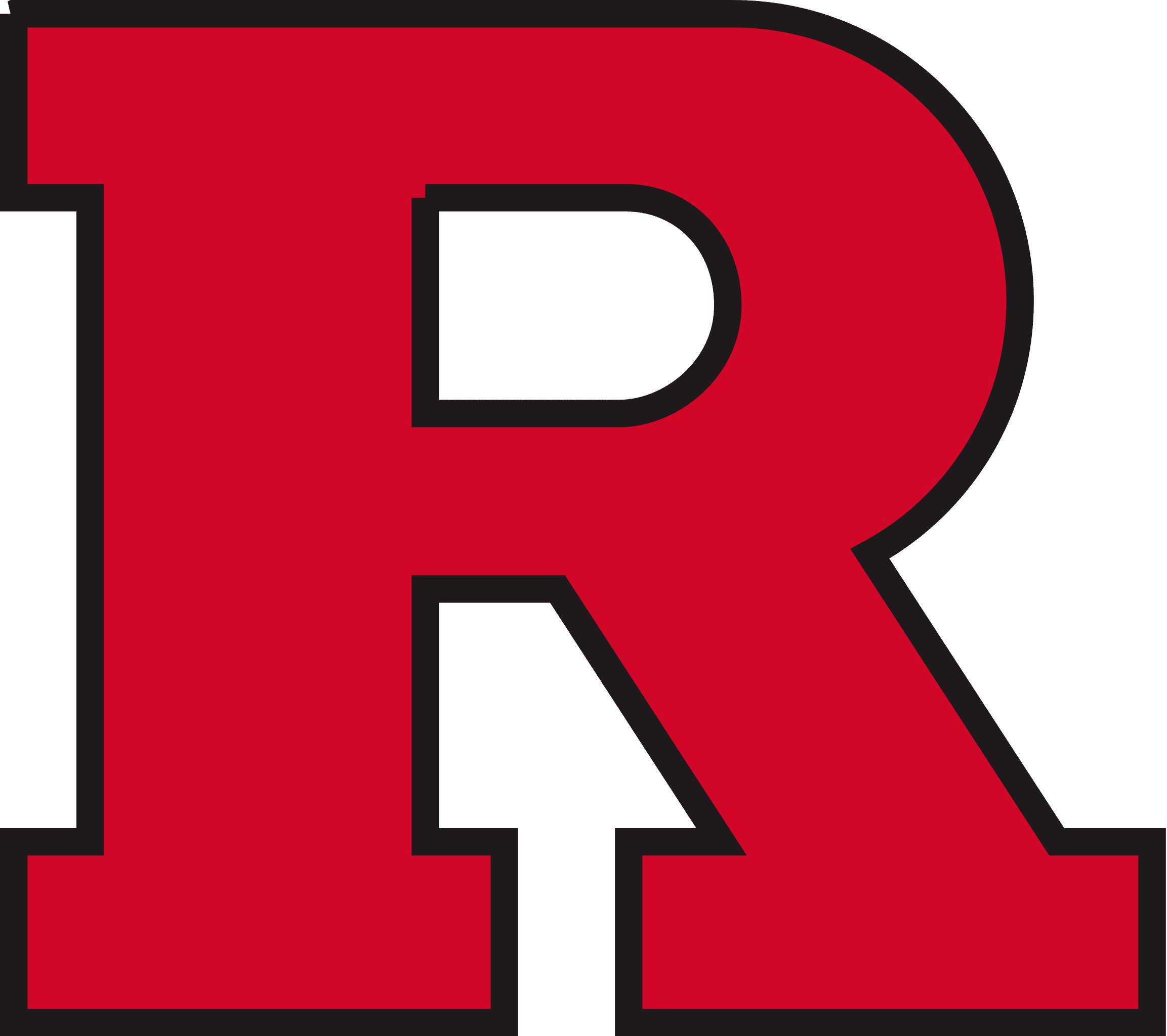}}
\newcommand{\gdmlogo}{\hspace{0.8pt}\includegraphics[height=0.8em]{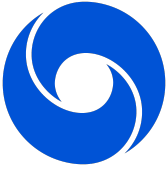}}

\newcommand{\ours}{{\texttt{Concorde}}\xspace}
\newcommand{\ourss}{{\small{{\texttt{Concorde}}}}\xspace}

\newcommand{\workload}[1]{{\fontfamily{qcr}\selectfont #1}}

\newcommand{\niparagraph}[1]{\smallskip\noindent\textbf{#1.}}
\newcommand{\ninparagraph}[1]{\smallskip\noindent\textbf{#1}}

\newcommand{\specialcell}[2][c]{%
  \begin{tabular}[#1]{@{}c@{}}#2\end{tabular}}
\newcommand*\circled[1]{\tikz[baseline=(char.base)]{
            \node[shape=circle,draw,inner sep=0.4pt] (char) {#1};}}

\newcommand{\egc}{e.g.,\ }
\newcommand{\iec}{i.e.,\ }

\newcommand{\ooo}{OoO\ }
\newcommand{\xx}[1]{\textsf{#1}\xspace}

\newcommand*\squared[1]{\tikz[baseline=(char.base)]{\node[shape=rectangle, fill=black, rounded corners=3pt, inner sep=1pt, minimum size=1em] (char) {\textcolor{white}{#1}};}}

\title[Concorde: Fast and Accurate CPU Performance Modeling with Compositional Analytical-ML Fusion]{Concorde: Fast and Accurate CPU Performance Modeling\\with Compositional Analytical-ML Fusion}

\author[A. Nasr-Esfahany et al.]{
\normalsize{Arash Nasr-Esfahany$^{\ast}$\textsuperscript{\nospacegooglelogo\mitlogo}},\,Mohammad Alizadeh$^{\ast}$\textsuperscript{\nospacegooglelogo\mitlogo},\,Victor Lee\textsuperscript{\googlelogo},\,Hanna Alam\textsuperscript{\googlelogo},\,Brett W. Coon\textsuperscript{\googlelogo}\\
David Culler\textsuperscript{\googlelogo},\,Vidushi Dadu\textsuperscript{\googlelogo},\,Martin Dixon\textsuperscript{\googlelogo},\,Henry M. Levy\textsuperscript{\googlelogo\uwlogo},\,Santosh Pandey$^{\ast}$\textsuperscript{\nospacegooglelogo\rutgerslogo}\\
Parthasarathy Ranganathan\textsuperscript{\googlelogo},\,Amir Yazdanbakhsh\textsuperscript{\gdmlogo}\\\vspace{2.5mm}
Google\textsuperscript{\googlelogo}\hspace{-2pt},\,MIT\textsuperscript{\mitlogo}\hspace{-2pt},\,University of Washington\textsuperscript{\uwlogo}\hspace{-2pt},\,Rutgers University\textsuperscript{\rutgerslogo}\hspace{-1pt},\,Google DeepMind\textsuperscript{\gdmlogo}\\\vspace{2.5mm}
\footnotesize{\texttt{{\{\href{mailto:arashne@mit.edu}{arashne},\,\href{mailto:alizadeh@mit.edu}{alizadeh}\}@mit.edu}}},\,\footnotesize{\texttt{\href{mailto:santosh.pandey@rutgers.edu}{santosh.pandey}@rutgers.edu}},\,\footnotesize{\texttt{\{\href{mailto:vwlee@google.com}{vwlee},\,\href{mailto:mailto:hannaalam@google.com}{hannaalam},\,\href{mailto:mailto:bwc@google.com}{bwc},\,\href{mailto:dculler@google.com}{dculler},\,\href{mailto:vidushid@google.com}{vidushid},\,\href{mailto:mgdixon@google.com}{mgdixon},\,\href{mailto:hanklevy@google.com}{hanklevy},\,\href{mailto:parthas@google.com}{parthas},\,\href{mailto:ayazdan@google.com}{ayazdan}\}@google.com}}
}

\begin{abstract}
Cycle-level simulators such as gem5 are widely used in microarchitecture design, but they are prohibitively slow for large-scale design space explorations.
We present \ours, a new methodology for learning fast and accurate performance models of microarchitectures.
Unlike existing simulators and learning approaches that emulate each instruction, \ours predicts the behavior of a program based on compact performance distributions that capture the impact of different microarchitectural components. 
It derives these performance distributions using simple analytical models that estimate bounds on performance induced by each microarchitectural component, providing a simple yet rich representation of a program's performance characteristics across a large space of microarchitectural parameters. 
Experiments show that \ours is more than five orders of magnitude faster than a reference cycle-level simulator, with  about 2\% average Cycles-Per-Instruction (CPI) prediction error across a range of SPEC, open-source, and proprietary benchmarks. 
This enables rapid design-space exploration and performance sensitivity analyses that are currently infeasible, e.g., in about an hour, we conducted a first-of-its-kind fine-grained performance attribution to different microarchitectural components across a diverse set of programs, requiring nearly 150 million CPI evaluations.
\end{abstract}
\maketitle
\def\thefootnote{$\ast$}\footnotetext{Work done at Google.}
\setcounter{footnote}{0}
\renewcommand{\thefootnote}{\arabic{footnote}}
\section{Introduction}
\label{sec:intro}

Microarchitecture simulators are a key tool in the computer architect's arsenal~\cite{skadron2003challenges,gem5,patel2011marss,austin2002simplescalar,carlson2011sniper,sanchez2013zsim,akram2019survey,elrabaa2017very,alves2015sinuca}.
From SimpleScalar~\cite{austin2002simplescalar} to gem5~\cite{gem5,gem5update}, simulators have enabled architects to explore new designs and optimize existing ones without the prohibitive costs of fabrication.
CPU simulation, in particular, has become increasingly important as hyperscale companies like Google (Axion)~\cite{axion}, Amazon (Graviton)~\cite{graviton}, Microsoft (Cobalt)~\cite{cobalt} increasingly invest in developing custom CPU architectures tailored to their specific workloads.

The landscape of CPU performance modeling is characterized by a critical tension between model accuracy and speed. 
This trade-off manifests in the various levels of abstraction employed by different performance models~\cite{gem5,genbrugge2010interval,analytical}.
At one end of the spectrum lie analytical models~\cite{analytical,abel2023facile}, which provide simplified mathematical representations of microarchitectural components and their interactions. 
Although they are fast, analytical models often lack the detailed modeling necessary to capture the dynamics of modern processors accurately. 
At the other end of the spectrum reside cycle-level simulators like gem5~\cite{gem5}, which can provide high-fidelity results by meticulously modeling every cycle of execution.  
However, this level of detail comes at a steep computational cost, becoming prohibitively slow for large-scale design space exploration~\cite{karandikar2018firesim, genbrugge2010interval, simnet}, programs with billions of instructions, or detailed sensitivity studies.

Recognizing the limitations of conventional methods, there has been growing interest in using machine learning (ML) to expedite CPU simulation~\cite{mendis2019ithemal,simnet,tao,perfvec}.
Rather than explicitly model every cycle and microarchitectural interaction, these methods learn an approximate model of the architecture's performance from a large corpus of data. 
A typical approach is to pose the problem as learning a function mapping a sequence of instructions to the target performance metrics. For example, recent work~\cite{tao,offline,simnet,perfvec} train sequence models (e.g., LSTMs~\cite{lstm} and Transformers~\cite{transformer}) on ground-truth data from a cycle-level simulator to predict metrics such as the program's Cycles Per Instruction (CPI).

These methods show promise in providing fast performance estimates with reasonable accuracy.
However, relying on black-box ML models operating on instruction sequences has several limitations.
First, the computational cost of these methods scales proportionally with the length of the instruction sequence, i.e. $\mathcal{O}(L)$ where $L$ is the instruction sequence length.
The $\mathcal{O}(L)$ complexity limits the potential speedup of these methods, e.g., to less than $10\times$ faster than cycle-level simulation with a single GPU~\cite{simnet, tao}. 
This speedup is mainly due to replacing the irregular computations of cycle-level simulation with accelerator-friendly neural network calculations~\cite{tao}. 
By contrast, analytical models can be several orders of magnitude faster than cycle-level simulation (and current ML approaches) because they fundamentally operate at a higher level of abstraction, i.e., mathematical expressions relating key statistics (e.g., instruction mix, cache behavior, branch misprediction rate, etc.) to performance. 

Second, existing ML approaches must learn all the dynamics impacting performance from raw instruction-level training data.
In many cases, this learning task is unnecessarily complex since it does not exploit the CPU performance modeling problem structure.
For example, TAO's Transformer model~\cite{tao} must learn the performance impact of register dependencies from per-instruction register information, even though there exist higher-level abstractions (e.g., instruction dependency graphs~\cite{mendis2019ithemal}) that concisely represent dependency behavior (\S\ref{sub:design:analytical}). 
By ignoring the problem structure, blackbox methods require a significant amount of training data to learn. For example, TAO trains on a dataset of 180 million instructions across four benchmarks and two microarchitectures~\cite{tao}, with further training required for each new microarchitecture. 

To address these challenges, we propose a novel approach to performance modeling---compositional analytical-ML fusion---where we decompose the task into multiple lightweight models that work together to progressively achieve high fidelity with low computational complexity. We demonstrate this approach in \ours, a CPU performance model that uses simple analytical models capturing the first-order effects of individual microarchitectural components, coupled with an ML model that captures complex higher-order effects (e.g., interactions of multiple microarchitectural components). 

\ours achieves constant-time $\mathcal{O}(1)$ inference complexity, independent of the length of the instruction stream, while maintaining high accuracy across diverse workloads and microarchitectures. Unlike existing ML methods that operate on instruction sequences, \ours predicts performance based on a compact set of {\em performance distributions}.
It trains a lightweight ML model\,---\,a shallow multi-layer perceptron (MLP)\,---\,to map these performance distributions to the target performance metric. We focus on modeling CPI in this paper as it directly reflects program performance, though in principle our techniques could be extended to other metrics. 
\ours's ML model generalizes across a large space of designs, specified via a set of parameters associated with different microarchitectural components (\S\ref{sec:design}).
Given the performance distributions for a program region (e.g., 1M instructions), predicting its CPI on any target microarchitecture is extremely fast; it requires only a single neural network evaluation, taking less than a millisecond.

\ours derives a program region's performance distributions through a two-step process: trace analysis and analytical modeling.
Trace analysis uses simple in-order cache and branch predictor simulators to extract information such as instruction dependencies, approximate execution latencies, and branch misprediction rate.
Next, analytical models estimate the bottleneck throughput imposed by each CPU component (e.g., fetch buffer, load queue, etc.) in isolation, assuming other CPU components have infinite capacity.
For each CPU component, \ours uses the distribution of its throughput bound over windows of a few hundred instructions as its performance feature.
For each memory configuration, \ours executes the per-component analytical models independently to precompute the set of performance distributions for all parameter values. 
The analytical models are lightweight, completing in 10s of milliseconds for a million instructions. Precomputing the performance distributions
is a one-time cost, enabling nearly instantaneous performance predictions across the entire parameter space.

\ours's unique division of labor between analytical and ML modeling simplifies both the analytical and ML models.
Since the analytical models are not directly used to predict performance, they are relatively easy to construct. 
Their main goal is to provide a first-cut estimate of the performance bounds associated with each microarchitectural component (akin to roofline analysis~\cite{roofline}), without the burden of quantifying the combined effect of multiple interacting components. 
The ML model, on the other hand, starts with features that correlate strongly with a program's performance, rather than raw instruction sequences. 
Its task is to capture the higher-order effects ignored by the analytical models, such as the impact of multiple interacting bottlenecks. 
The net result is a method that is as fast as analytical models while achieving high accuracy.

\ours enables large-scale analyses that are deemed impractical with conventional methods. 
As one use case, we consider the problem of fine-grained performance attribution to different microarchitectural components: 
{\em What is the relative contribution of different microarchitectural components to the predicted performance of a target architecture?}
We present a novel technique for answering this question using only a performance model relating microarchitectural parameters to performance.
Our technique applies the concept of Shapley value~\cite{shapley} from cooperative game theory to provide a fair and rigorous attribution of performance to individual  components. 
The method improves upon standard parameter ablation studies and may be of interest in other use cases beyond \ours. 

We present a concrete realization of \ours, designed to approximate the behavior of a proprietary gem-5 based cycle-level trace-driven CPU simulator. 
We train \ours on a dataset of 1~million random program regions and architectures, to predict the impact of 20 parameters spanning frontend, backend, and memory (totaling $2.2\times 10^{23}$ parameter combinations). 
The program regions are sampled from a diverse set of SPEC2017~\cite{spec2017}, open-source, and proprietary benchmarks. The key findings of our evaluation are: 
\begin{itemize}[leftmargin=*]
    \item \ours's average CPI prediction error is within 2\% of the ground-truth cycle-level simulator for unseen (random) program regions and architectures, with only 2.5\% of samples exceeding a 10\% prediction error. Ignoring the one-time cost of analytical modeling, \ours is five orders of magnitude faster for predicting the performance of 1M-instruction regions. For long 1B~instruction programs, \ours accurately estimates performance (average error $\sim$3.2\%) based on randomly-sampled program regions, seven-orders of magnitude faster than cycle-level simulation. 
   
    \item In predicting CPI for a realistic core model (based on ARM N1~\cite{arm-n1}), \ours is more accurate than TAO~\cite{tao}, the state-of-the-art sequence-based ML performance model, trained specifically for the same core configuration. It achieves an average prediction error of $3.5\%$, compared to $7.8\%$ for TAO.
   
    \item For a 1M-instruction region, precomputing all the performance distributions takes the CPU time equivalent of 7 to 107 cycle-level simulations, depending on the granularity of parameter sweeps. These performance features enable rapid performance predictions for  $1.8\times10^{18}$ to $2.2 \times 10^{23}$ parameter combinations.

    \item \ours enables a first of its kind, large-scale, fine-grained performance attribution to components of a core based on ARM N1, across a diverse set of programs using our Shapley value technique. This large-scale analyses requires more than $143$M CPI evaluations, but takes only about one hour with \ours. 
\end{itemize}
\section{Motivation and Insights}
\label{sec:motivation}

Consider a cycle-level simulator like gem5 as implementing a function that maps an input program and microarchitecture configuration to a performance metric such as CPI.
Formally, $y = f(\vec{\mathbf{x}}, \vec{\mathbf{p}})$, where $\vec{\mathbf{x}} \triangleq (x_1, \ldots, x_L)$ denotes the input program comprising $L$ instructions, $\vec{\mathbf{p}} \triangleq (p_1, \ldots, p_d)$ the parameters specifying the microarchitecture, and $y$ the CPI achieved by program $\vec{\mathbf{x}}$ on microarchitecture $\vec{\mathbf{p}}$.
Our goal is to learn a fast and accurate approximation of the function $f$ from training examples derived from cycle-level simulations.

Supervised learning provides the de facto framework for learning a function from input-output examples.
However, a critical design decision involves how to best represent the learning problem, including the selection of representative features and an appropriate model architecture.
Several recent efforts~\cite{simnet, tao, perfvec} represent the function $f$ using sequence-based models, such as LSTMs~\cite{lstm} and Transformers~\cite{transformer}, operating on raw or minimally processed instruction sequences.
As discussed in \S\ref{sec:intro}, these blackbox sequence models inherently limit scalability and increase the complexity of the learning task.
Our key insight is a novel decomposition of the function $f$, comprising of two key stages. First, an analytical stage uses simple per-component models to extract compact performance features capturing the overall performance characteristics of the program. Second, a lightweight ML model predicts the target CPI metric efficiently based on these performance features.

\niparagraph{Deriving compact performance features} 
The foundation of our analytical stage is to characterize the bottleneck throughput imposed by each CPU resource\footnote{
For simplicity, this section focuses on CPU parameters. \ours handles a few parameters such as cache sizes differently (\S\ref{sec:design}).} individually, under the simplifying assumption that all other CPU components operate with unlimited capacity.
For instance, to analyze the impact of the reorder buffer (ROB) size, we evaluate the program's throughput in a hypothetical system constrained only by the ROB size and instruction dependencies (i.e., a perfect frontend with no backend resource bottlenecks other than the limited ROB size).  
Focusing on one resource at a time enables relatively straightforward analyses (see~\S\ref{sub:design:analytical} for examples).
Formally, given a program $\vec{\mathbf{x}}$, we compute the bottleneck throughput $z_i = A_i(\vec{\mathbf{x}},p_i)$ for each CPU component, where $A_i(\cdot,\cdot)$ is the analytical model for the $i^\text{th}$ component, parameterized by $p_i$. 
To capture program phase changes, we calculate this throughput over small windows of consecutive instructions (e.g., a few hundred instructions).

\begin{figure}
\centerline{\includegraphics[width=\linewidth]{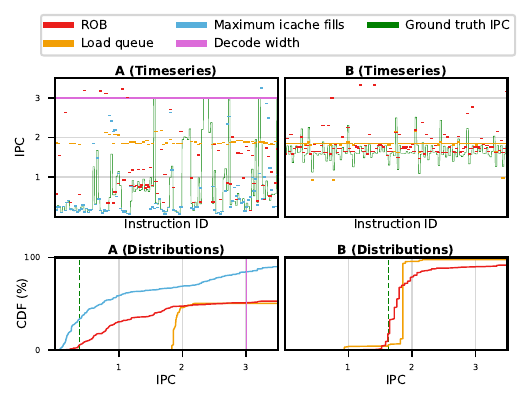}}
\vspace{-3mm}
\caption{\small Per-resource analytical modeling produces a rich performance characterization of a program.}
\vspace{-0.3cm}
\label{fig:analytical}
\end{figure}

\Cref{fig:analytical} shows an illustrative example of these throughput calculations for four microarchitectural parameters (ROB size, Load queue size, maximum I-cache fills, and decode width) on two programs.
The top plots display the timeseries of the throughput bounds derived by our analytical model for each parameter (details in \S\ref{subsub:design:per-resource-thr}) across 400-instruction windows, and the ground truth Instructions per Cycle (IPC) for the same windows.
For both programs, the throughput bound timeseries explain the IPC trends well.
For example, for program~A, initially the IPC (green line) aligns with the maximum I-cache fills bound (cyan segments); subsequently, the IPC is around the smaller of the ROB, decode width, and maximum I-cache fills bounds in most instruction windows. Similarly, for program B, the bounds for ROB and Load queue overlap with the IPC. The maximum I-cache fills and decode width throughput bounds are much higher for program B (not shown in the figure).

Although the minimum of the per resource throughput bounds provides an estimate of IPC, it is not accurate.
As shown in \Cref{fig:analytical}, despite their overall correlation, the IPC frequently deviates from the exact minimum bound.
This is not surprising. 
The analytical models make simplifying approximations, including ignoring interactions between multiple resource bottlenecks. 
In reality, resource bottlenecks can overlap, resulting in a net IPC lower than any individual bound.
Nonetheless, the per-resource analysis provides informative features for predicting performance, capturing key first-order effects while leaving it to the ML model to capture higher-order effects.

The last step of deriving compact performance features (independent of program length $L$) is converting throughput timeseries into {\em distributions}, as depicted in the bottom plots in \Cref{fig:analytical}. 
We encode these distributions using a fixed set of percentiles from their Cumulative Distribution Functions (CDF).
Converting timeseries to CDFs is inherently lossy (e.g., joint behaviors across timeseries are not retained). However, as \Cref{fig:analytical} shows, the CDFs are still informative for predicting IPC. In particular, the IPC (vertical dashed line)  aligns well with the lower percentiles of the smaller throughput bounds (e.g., Maximum I-cache fills and ROB for program A)\,---\,an implication of the IPC's proximity to the minimum throughput bound in most instruction windows. As our experimental results will show, a simple ML model can learn to accurately map these CDFs to IPC.

\begin{figure}
\centerline{\includegraphics[width=\linewidth]{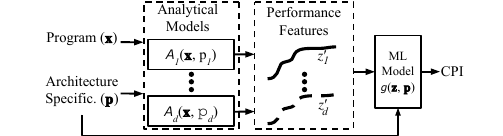}}
\caption{
\ours's compositional analytical-ML structure}
\vspace{-3mm}
\label{fig:oursio}
\end{figure}

\niparagraph{\ours's compositional analytical-ML structure}
\Cref{fig:oursio} illustrates the two-stage structure of \ours.
To predict the performance of program 
$\vec{\mathbf{x}}$ on a given microarchitecture $\vec{\mathbf{p}}$, \ours first uses per-component analytical models to derive performance features $\vec{\mathbf{z}} \triangleq ({z'_1}, \ldots, {z'_d})$, where ${z}'_i$ represents the distribution of the throughput bound for parameter $p_i$. 
These features, along with the list of parameters, are then passed to a lightweight ML model $\hat{y} = g(\vec{\mathbf{z}}, \vec{\mathbf{p}})$ to predict the CPI.

An important consequence of modeling each component (parameter) separately in the analytical stage is the ability to {\em precompute} the performance features for a program $(\vec{\mathbf{x}})$ across the entire microarchitectural design space. 
In particular, our approach eliminates the need to evaluate the Cartesian product of all parameters, which would require exponential time and space.
Instead, \ours sweeps the range of each CPU parameter (once or per memory configuration depending on the parameter), precomputing
the feature set $\{A_i(\vec{\mathbf{x}}, p_i) | \forall p_i, \forall i\}$.
To predict the performance of $\vec{\mathbf{x}}$ on a specific microarchitecture $\vec{\mathbf{p}}$, \ours retrieves the pertinent precomputed features corresponding to $p_1,\ldots,p_d$ and evaluates the ML model $g(\vec{\mathbf{z}},\vec{\mathbf{p}})$. 
\section{\ours Design}
\label{sec:design}
We present a concrete realization of \ours designed to approximate a proprietary gem5-based cycle-level trace-driven simulator. Inevitably, some aspects of \ours (esp., analytical models) depend on the specifics of the reference architecture. We detail the design for our in-house cycle-level simulator while emphasizing concepts that we believe apply broadly to CPU modeling.

\begin{figure}
    \centering
    \includegraphics[width=\linewidth]{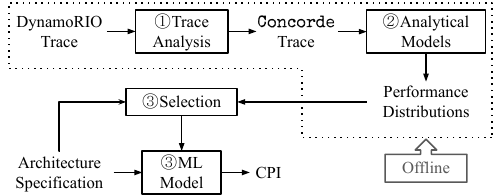}
    \caption{\small Design overview}
    \vspace{-3mm}
    \label{fig:design}
\end{figure}

Our cycle-level simulator processes program traces captured by DynamoRIO~\cite{dynamoRIO} and features a generic parameterized Out-of-Order (OoO) core model similar to gem5's O3 CPU model~\cite{gem5-O3-cpu}.
The architecture consists of fetch, decode, and rename stages in the frontend; issue, execute, and commit stages in the backend; and uses Ruby~\cite{ruby} for modeling the memory system.\footnote{
We use a fixed LLC size of 4MB, and a cache replacement policy similar to gem5's TreePLURP, a pseudo-Least Recently Used (PLRU) replacement policy.
Our cache allocation policy is the same as gem5's standard allocation policy, always allocating lines on reads and writebacks.
A cache line is not allocated on sequential access (for L2 and LLC) or a unique read for LLC.
We use writeback for all L1i, L1d, L2, and LLC.
We use memory BW of 37GB/s with latency of 90ns, and do not model memory channels.
}
We focus on modeling the impact of 20 key design parameters on CPI, as summarized in \Cref{tab:param-space}, though our approach can be extended to other design parameters.

\Cref{fig:design} outlines \ours's key design elements: \circled{1} \textbf{Trace analysis} which augments the input DynamoRIO~\cite{dynamoRIO} trace with information needed for \ours (\S\ref{sub:design:ta}); \circled{2} \textbf{Per-resource analytical models} which transform the processed trace into performance distributions (\S\ref{sub:design:analytical}); and \circled{3} \textbf{Lightweight ML model} which predicts the CPI based on performance distributions (\S\ref{sub:design:ml}).
The first two stages perform a one-time, offline computation for a given DynamoRIO trace. 
At simulation time, \ours supplies the precomputed performance distributions and target microarchitecture's design parameters to the ML model, enabling nearly instantaneous CPI predictions.

\subsection{Trace Analysis}
\label{sub:design:ta}
The raw input trace to \ours is captured using DynamoRIO's drmemtrace client~\cite{drmemtrace}, which provides detailed instruction and data access information for the target program.
This trace is then processed into a \ours Trace, which includes per-instruction information needed for our analytical models. 
We categorize this information into microarchitecture independent and microarchitecture dependent features, as detailed below.

\niparagraph{Microarchitecture independent}
This category includes data derived directly from the DynamoRIO trace:
\emph{(i)~Instruction dependencies}, including both register and memory dependencies,
\emph{(ii)~Program counters (PC)}  for all instructions,
\emph{(iii)~Data cache lines} for Load instructions,
\emph{(iv)~Instruction cache lines} for all instructions, 
\emph{(v)~Instruction Synchronization Barriers (ISB)}, and
\emph{(vi)~Branch types} (Direct unconditional, Direct conditional, and Indirect branches) for branch instructions.

\niparagraph{Microarchitecture dependent}
\emph{(i)~Execution latency:}
Our analytical models require an estimate of the execution latency for each instruction. For non-memory instructions, we estimate the latency based on the opcode and corresponding execution unit (e.g., 3 cycles for integer ALU operations). Store instructions also incur a fixed, known latency, as the architecture uses write-back (with store forwarding). Load instructions, however, have variable latency depending on the cache level. To estimate their latency, we perform a simple in-order cache simulation (per memory configuration) to determine the cache level for each Load. We then map each cache level to a constant latency (e.g., L1 $\rightarrow$ 4 cycles, L2 $\rightarrow$ 10 cycles, LLC~$\rightarrow$ 30 cycles, RAM $\rightarrow$ 200 cycles).
\emph{(ii)~I-cache latency:}
To model the fetch stage, our analytical models need an estimate of I-cache access times, which we obtain by performing a simple in-order I-cache simulation (per memory configuration).
\emph{(iii)~Branch misprediction rate}, which we obtain by simulating the target branch prediction algorithm on the DynamoRIO trace. Our implementation supports two branch predictors: {\em Simple}, a branch predictor that mispredicts randomly with a pre-specified misprediction rate, and {\em TAGE}~\cite{andre2006case,seznec2011new}.

\begin{table}
\scriptsize
    \centering
    \caption{\small Large space of design parameters}
    \vspace{-3mm}
    \label{tab:param-space}
    \begin{tabular}{c c c}
        \toprule
        \textbf{Parameter} & \textbf{Value Range} & \textbf{ARM N1 value}\\
        \midrule
            ROB size & $1, 2, 3, \ldots, 1024$ & $128$\\
            Commit width & $1, 2, 3, \ldots, 12$ & $8$\\
            Load queue size & $1, 2, 3, \ldots, 256$ & $12$\\
            Store queue size & $1, 2, 3, \ldots, 256$ & $18$\\
            ALU issue width & $1, 2, 3, \ldots, 8$ & $3$\\
            Floating-point issue width & $1, 2, 3, \ldots, 8$ & $2$\\
            Load-store issue width & $1, 2, 3, \ldots, 8$ & $2$\\
            Number of load-store pipes & $1, 2, 3, \ldots, 8$ & $2$\\
            Number of load pipes & $0, 1, 2, \ldots, 8$ & $0$\\
        \midrule
            Fetch width & $1, 2, 3, \ldots, 12$ & $4$\\
            Decode width & $1, 2, 3, \ldots, 12$ & $4$\\
            Rename width & $1, 2, 3, \ldots, 12$ & $4$\\
            Number of fetch buffers & $1, 2, 3, \ldots, 8$ & $1$\\
            Maximum I-cache fills & $1, 2, 3, \ldots, 32$ & $8$\\
            Branch predictor & Simple, TAGE & TAGE\\
            Percent misprediction for \textit{Simple} BP & $0, 1, 2, \ldots, 100$ & $-$\\
        \midrule
            L1d cache size (kB) & $16, 32, 64, 128, 256$ & $64$\\
            L1i cache size (kB) & $16, 32, 64, 128, 256$ & $64$\\
            L2 cache size (kB) & $512, 1024, 2048, 4096$ & $1024$\\
            L1d stride prefetcher degree & $0\text{ (OFF)}, 4\text{ (ON)}$ & $0$(OFF)\\
        \bottomrule
    \end{tabular}
    \vspace{-3mm}
\end{table}

\niparagraph{Improving memory modeling}
\label{sub:design:memory-state-machine}
The execution times assigned to Load instructions by the above procedure can be highly inaccurate in some cases, leading \ours's analytical models astray. As we will see, \ours's ML model can overcome many errors in the analytical model. However, in extreme cases at the tail, \ours's accuracy is affected by discrepancies between the results of trace analysis and the program's actual behavior (\S\ref{sub:eval:tail-analysis}).

The key challenge with analyzing Load instructions is that their execution times can change depending on the time and order in which they are issued. Of course, a simple in-order cache simulation cannot capture timing-dependent effects. However, we now discuss a refinement atop the basic cache simulation that addresses two large sources of errors in estimating Load execution times. Our approach is built on two principles for accounting for the effects of conflicting cache lines and instruction order {\em without} running detailed timing simulations.

Consider two Load instructions accessing the same cache line, with the data not present in cache.
In the in-order cache simulation, the first Load is labeled as a main memory access (200 cycles), while the second Load is labeled as an L1 hit (4 cycles).
Now suppose these Loads are issued at around the same time in the actual \ooo core, e.g., first Load at cycle~0 and second Load at cycle~1. 
Na\"ively using the cache simulation results, we might conclude that the first Load completes at cycle~200 and the second Load at cycle~5. 
But, in reality, both Loads will complete after 200 cycles because the second Load must wait for the first Load to fetch the data from main memory into L1 cache.   
This example motivates our first principle: \emph{the response cycle for consecutive Loads accessing the same cache line is non-decreasing.}

Next, consider the same scenario, but with the two Loads issued in reverse order in the \ooo core (e.g., due to a register dependency). 
With this reversed order, the second Load (issued first) becomes a main memory access, and the first Load (issued second) becomes an L1 hit. Thus, our second principle: \emph{the access levels of Loads with the same cache line is determined by their issue order, not the instruction order (used in cache simulation).}
\begin{algorithm}[t]
\caption{A trace-driven state machine for memory}
\label{alg:mem-state-machine}
\begin{algorithmic}
\footnotesize
\ForAll{$cache\_line$} \Comment{\texttt{State variable initialization}}
    \State exec\_times$[cache\_line] \gets$ Execution times of load instructions accessing
    \State \textcolor{white}{exec\_times$[cache\_line] \gets$}$cache\_line$ from in-order cache simulation
    \State access\_counters$[cache\_line] \gets 0$  \Comment{\texttt{Number of accesses}}
    \State last\_req\_cycles$[cache\_line] \gets 0$ \Comment{\texttt{Cycle of last request}}
    \State last\_resp\_cycles$[cache\_line] \gets 0$ \Comment{\texttt{Cycle of last response}}

\EndFor
\State
\Function{RespCycle}{$req\_cycle, instr$} 
    \State $cache\_line \gets instr.cache\_line$ \\\Comment{\texttt{$req\_cycle$ must be non-decreasing for requests to the same cache line}}
    \State {\bf Assert} $req\_cycle \geq \text{last\_req\_cycle}[cache\_line]$  
    \If{$\text{is\_load}(instr)$} \Comment{\texttt{Adjustment for load instructions only}}
        \State prev\_resp\_cycle $\gets$ last\_resp\_cycles$[cache\_line]$
        \State access\_number $\gets$ access\_counters$[cache\_line]$
        \State exec\_time $\gets$ exec\_times$[cache\_line][\text{access\_number}]$ 
        \State resp\_cycle $\gets$ $\max(req\_cycle + \text{exec\_time}, \text{prev\_resp\_cycle})$
        \State last\_resp\_cycles$[cache\_line]\gets$ resp\_cycle
        \State access\_counters$[cache\_line]++$
    \Else \Comment{\texttt{Nothing special for non-load instructions}}
        \State resp\_cycle $\gets$ $req\_cycle$ $+$ estimated execution time of $instr$
    \EndIf
    \State\textbf{return} resp\_cycle
\EndFunction
\end{algorithmic}
\end{algorithm}
We incorporate these principles into a trace-driven state machine for memory (\Cref{alg:mem-state-machine}).
The function  \textsc{RespCycle} returns the response cycle (execution completion cycle) for an instruction issued at cycle $req\_cycle$. For non-Load instructions, it simply uses the execution time estimated by the standard procedure described earlier. For Loads, however, it adjusts the execution time to account for their cache line and issue times. We use this memory model in the analytical models of ROB and Load queue, which are sensitive to Load execution latencies (\S\ref{sub:design:analytical}). The memory model is fast and does not materially increase the cost of analytical modeling. 

\subsection{Analytical Models}
\label{sub:design:analytical}
As discussed in \S\ref{sec:motivation}, \ours's primary features are a set of throughput distributions associated with each potential microarchitectural resource bottleneck. We describe the derivation of these distributions for various resources in \S\ref{subsub:design:per-resource-thr}. We then discuss a few auxiliary features in \S\ref{sub:design:additional-feat} that capture nuances not covered by the primary features, further improving the ML model's accuracy. 

The bulk of \ours's design effort has gone into analytical modeling. Before delving into details, we highlight a few lessons from our experience. Our guiding principle has been to capture the \emph{performance trends} imposed by a microarchitectural bottleneck, without being overly concerned with precision. As our results will show (\S\ref{sub:eval:ablation}), the ML model serves as a powerful backstop that can mask signifcant errors in the analytical model. Thus, we have generally avoided undue complexity (admittedly a subjective metric!) to improve the analytical model's accuracy. Our decision to simply analyze each resource in isolation (\S\ref{sec:motivation}) is the clearest example of this philosophy.

Isolated per-resource throughput analysis is similar to traditional roofline analysis~\cite{roofline}, but we perform it at an unusually fine granularity to analyze the impact of low-level resources (e.g., an issue queue, fetch buffers, etc.) on small windows (e.g., few 100s) of instructions. The details of such analyses depend on the design, but we have found three types of models to be useful: (i) closed-form mathematical expressions, (ii) dynamical system equations, (iii) simple discrete-event simulations of a single component. We provide examples of these methods below.

\subsubsection{Per-Resource Throughput Analysis}
\label{subsub:design:per-resource-thr}
We calculate the throughput of each CPU resource over fixed windows of $k$ consecutive instructions, assuming no other CPU component is bottlenecked. The parameter $k$ should be small enough to observe phase changes in the program's behavior, but not so small that throughput fluctuates wildly due to bursty instruction processing (e.g., a few instructions). We have found that any value of $k$ in the order of the ROB size, typically a few hundred instructions, works well.

Given a program region (e.g., 100K-1M instructions), \ours divides it into consecutive $k$-instruction windows and calculates the throughput bound for each window, per CPU resource and parameter value (\Cref{tab:param-space}). \ours converts all throughput bound timeseries into distributions (CDFs) to arrive at the set of performance distributions for the entire microarchitectural design space.

Memory parameters (L1i/d, L2, L1d prefetcher degree) do not have separate throughput features; they affect the instruction execution latency and I-cache latency estimates (\S\ref{sub:design:ta}) used in CPU resource analyses. Specifically, the throughput computations for ROB, and Load/Store queues rely on instruction execution latencies. \ours performs throughput calculations for these resources per L1d/L2/prefetch configuration using the corresponding execution latency values in the \ours trace. Similarly, the I-cache fills throughput calculations are performed per L1i/L2 cache size. 

\niparagraph{ROB}
The ROB is the most complex component to model, encapsulating out-of-order execution constrained by instruction dependencies and in-order commit behavior. 
For an instruction $i$, we define Dep$(i)$ as its immediate (register and memory) dependencies obtained via trace analysis (\S\ref{sub:design:ta}), $a_i$ as its arrival cycle to the ROB, $s_i$ as its execution start cycle, $f_i$ as its execution finish cycle, and $c_i$ as its commit cycle.
We calculate the throughput induced by a ROB of size {\texttt{ROB}} using the following instruction-level dynamical system:
\begin{align}
    a_i &= c_{i-\text{\tt ROB}}, \label{eqn:rob:size}\\
    s_i &= \max\Bigl(a_i, \max\bigl\{f_d|d\in \text{Dep}(i)\bigr\}\Bigr),  \label{eqn:rob:dep}\\
    f_i &= \text{\textsc{RespCycle}}\Bigl(s_i, \text{instr}_i\Bigr), \label{eqn:rob:mem}\\
    c_i &= \max(f_i, c_{i-1}), \label{eqn:rob:commit}
\end{align}
for $i \geq 1$, where $c_i = 0$ for $i \leq 0$ by convention. 
\Cref{eqn:rob:size} enforces the size constraint of the ROB. 
\Cref{eqn:rob:dep} accounts for the instruction dependency constraints.
\Cref{eqn:rob:mem} uses the function shown in Algorithm~\ref{alg:mem-state-machine} (\S\ref{sub:design:memory-state-machine}) to determine the finish time of each instruction.\footnote{We execute \Cref{eqn:rob:mem} in order of instruction start times $s_i$ to satisfy \Cref{alg:mem-state-machine}'s requirement for non-decreasing request cycles.}
\Cref{eqn:rob:commit} models the in-order commit constraint.
Finally, the throughput for the $j^\text{th}$ window of $k$ instructions is calculated as:
\begin{equation}
    \textit{thr}^j_\textit{ROB} = \frac{k}{c_{kj} - c_{k(j-1)}}.
\label{eqn:rob:window-thr}
\end{equation}

\niparagraph{Load/Store queue}
The Load and Store queues bound the number of issued memory instructions that have yet to be committed (in order).
We briefly discuss the Load queue model (Store queue is analogous). It is identical to the ROB model, with two differences: (i) the calculations are performed exclusively for Load instructions,  (ii)  there are no dependency constraints: a Load is eligible to start as soon as it obtains a slot in the queue.  After computing the commit cycle for each Load, we derive the throughput for each $k$-instruction window similarly to Equation~\eqref{eqn:rob:window-thr}. In these calculations, non-Load operations are assumed to be free and incur no additional latency.

\niparagraph{Static bandwidth resources}
These resources impose limits on the number of instructions (of a certain type) that can be serviced in a single clock cycle.
For example, Commit, Fetch, Decode, and Rename widths constrain the throughput of all instructions. The throughput bound imposed by these resources is trivially their respective width. 
In contrast, issue queues restrict the throughput for a specific group of instructions, e.g., ALU, Floating-point, and Load-Store issue widths in our reference architecture. To compute the throughput bound imposed by such resources, we compute the processing time of the instructions that are constrained by that resource and assume non-affected instructions incur no additional latency. For instance, the throughput bound induced by the ALU issue width in the $j$-th window of $k$ consecutive instructions is given by:
\begin{equation}
    \textit{thr}_\textit{ALU}^j = \frac{k}{n^j_{\text{ALU}}}\times\text{ALU issue width},
\end{equation}
where $n^j_{\text{ALU}}$ is the number of ALU instructions in window $j$.

\niparagraph{Dynamic constraints}
Some resources impose constraints on a dynamic set of instructions determined at runtime based on the microarchitectural state. Analyzing such resources is more challenging. Two strategies that we have found to be helpful are to use simplified performance bounds or basic discrete-event simulation. We briefly discuss these strategies using two examples.

\noindentparagraph{Load/Load-Store Pipes.}
Finite Load and Load-Store pipes limit the number of memory instructions that can be issued per cycle. 
Store instructions exclusively use Load-Store pipes, while Load instructions can utilize both Load pipes and Load-Store pipes. 
The allocation of instructions to these pipes depends on dynamic microarchitectural state, e.g., the precise order that memory instructions become eligible for issue and the exact pipes available at the time of each issue. Rather than model such complex dynamics, we derive simple upper and lower bounds on the throughput. Let $n_{Load}$ and $n_{Store}$ denote the number of Load and Store instructions in a  $k$-instruction window, $LSP$ the number of Load-Store pipes, and $LP$ the number of Load pipes. The worst-case allocation of pipes is to issue Loads first using all available pipes, and only then begin issuing Stores using the Load-Store pipes. This allocation leaves the Load pipes idle while Stores are being issued. It results in the maximum total processing time: $T_{max} = n_{Load}/(LSP + LP) + n_{Store}/LSP$, and thus a lower-bound on the throughput of the pipes component: $thr_{lower} = k/T_{max}$. 

The best-case allocation is to grant Stores exclusive access to Load-Store pipes while concurrently using Load pipes to issue Loads. Once all Stores are issued, the Load Store pipes are allocated to the remaining Loads. Analogous to the lower bound, we can derive an upper bound on the throughput $thr_{upper}$ based on this allocation (details omitted for brevity). We summarize these bounds using the distribution of $thr_{lower}$ and $thr_{upper}$ over all instruction windows. 

\noindentparagraph{I-cache fills and fetch buffers.}
We model these resources using simple instruction-level simulations. Here, we focus on I-cache fills for brevity.
The maximum I-cache fills restricts the number of in-flight I-cache requests at any given time. This is a dynamic constraint, because whether an instruction generates a new I-cache request depends on the set of in-flight I-cache requests when it reaches the fetch target queue. Specifically, new requests are issued only for cache lines that are not already in-flight. We estimate the throughput constraint imposed by the maximum I-cache fills using a basic simulation of I-cache requests. This simulation assumes a backlog of instructions waiting to be fetched, restricted only by the availability of I-cache fill slots. Instructions are considered in order, and if they need to send an I-cache request, they send it as soon as an I-cache fill slot becomes available. We record the I-cache response cycle for each instruction in the simulation, and use it to calculate the throughput for each window of $k$ consecutive instructions similarly to \Cref{eqn:rob:window-thr}.

\subsubsection{Auxiliary Features}
\label{sub:design:additional-feat}
In addition to the primary features described above, we describe a few auxiliary features that capture nuances not covered by per-resource throughput analysis. We evaluate the impact of these auxiliary features in \S\ref{subsub:eval:additional-features}.

\niparagraph{Pipeline stalls}
Unlike resource constraints, modeling the effects of pipeline stalls caused by branch mispredictions and ISB instructions as an isolated component is not meaningful.
The impact of stalls on performance depends on factors beyond the fetch stage, for instance, the inherent instruction-level parallelism (ILP) of the program, how long it takes to drain the pipeline, and how quickly the stall is resolved~\cite{mispred-characterization}. Rather than try to model these complex dynamics analytically, we incorporate two simple groups of features to assist the ML model with predicting the impact of pipeline stalls. First, we provide basic information about the extent of stalls: (i) the distribution of the number of ISBs in our windows of $k$ consecutive instructions; (ii) the distribution of the count of the three branch types (\S\ref{sub:design:ta}) per instruction window, (iii) the overall branch misprediction rate obtained from trace analysis. Additionally, we provide the overall throughput calculated by our analytical ROB model (\S\ref{subsub:design:per-resource-thr})  for varying ROB sizes, $\texttt{ROB} \in \{1, 2, 4, 8, \ldots, 1024\}$. The intuition behind this feature is that pipeline stalls effectively reduce the average occupancy of the ROB, lowering the backend throughput of the CPU pipeline. Therefore, the ROB model's estimate of how throughput varies versus ROB size can provide valuable context for how sensitive a program's performance is to pipeline stalls. 

\begin{table}
\footnotesize
    \centering
    \caption{\small Workload space with $\mathbf{5486}$B instructions from $\mathbf{29}$ programs}
    \vspace{-3mm}
    \begin{adjustbox}{width=0.8\linewidth}
    \begin{tabular}{c c c c}
    \toprule
        \textbf{Type} & \textbf{Name} & \textbf{Traces} & \textbf{Instructions (M)} \\
    \midrule
        \multirow{13}{*}{Proprietary} & \workload{Compression \textbf{(P1)}} & $4$ & $1845$\\
                 & \workload{Search1 \textbf{(P2)}} & $168$ & $17854$\\
                 & \workload{Search4 \textbf{(P3)}} & $170$ & $23188$\\
                 & \workload{Disk \textbf{(P4)}} & $168$ & $23441$\\
                 & \workload{Video \textbf{(P5)}} & $268$ & $26981$\\
                 & \workload{NoSQL Database1 \textbf{(P6)}} & $168$ & $30283$\\
                 & \workload{Search2 \textbf{(P7)}} & $84$ & $52989$\\
                 & \workload{MapReduce1 \textbf{(P8)}} & $84$ & $56677$\\
                 & \workload{Search3 \textbf{(P9)}} & $1334$ & $69277$\\
                 & \workload{Logs \textbf{(P10)}} & $191$ & $75845$\\
                 & \workload{NoSQL Database2 \textbf{(P11)}} & $84$ & $91274$\\
                 & \workload{MapReduce2 \textbf{(P12)}} & $84$ & $104750$\\
                 & \workload{Query Engine\&Database \textbf{(P13)}} & $790$ & $1195128$\\
    \midrule
        \multirow{2}{*}{\specialcell{Cloud\\Benchmark}} & \workload{Memcached \textbf{(C1)}} & $8$ & $2791$\\
                        & \workload{MySQL \textbf{(C2)}} & $84$ & $9283$\\
    \midrule
        \multirow{4}{*}{\specialcell{Open\\Benchmark}} & \workload{Dhrystone \textbf{(O1)}} & $1$ & $174$\\
                       & \workload{CoreMark \textbf{(O2)}} & $1$ & $335$\\
                       & \workload{MMU \textbf{(O3)}} & $132$ & $18475$\\
                       & \workload{CPUtest \textbf{(O4)}} & $138$ & $95215$\\
    \midrule
        \multirow{10}{*}{SPEC2017} & \workload{505.mcf\_r \textbf{(S1)}} & $19$ & $197232$\\
                 & \workload{520.omnetpp\_r \textbf{(S2)}} & $20$ & $214749$\\
                 & \workload{523.xalancbmk\_r \textbf{(S3)}} & $20$ & $214749$\\
                 & \workload{541.leela\_r \textbf{(S4)}} & $20$ & $214749$\\
                 & \workload{548.exchange2\_r \textbf{(S5)}} & $20$ & $214749$\\
                 & \workload{531.deepsjeng\_r \textbf{(S6)}} & $20$ & $214749$\\
                 & \workload{557.xz\_r \textbf{(S7)}} & $38$ & $408022$\\
                 & \workload{500.perlbench\_r \textbf{(S8)}} & $41$ & $440235$\\
                 & \workload{525.x264\_r \textbf{(S9)}} & $44$ & $472447$\\
                 & \workload{502.gcc\_r \textbf{(S10)}} & $94$ & $999282$\\
    \bottomrule
    \end{tabular}
    \end{adjustbox}
    \label{tab:workload-space}
\end{table}

\niparagraph{Latency distributions}
We augment our primary throughput based features from \S\ref{sub:design:analytical} with three instruction-level latency distributions collected from the ROB model. Specifically, we provide the distribution of the time that instructions spend in the issue ($s_i - a_i$), execution ($f_i - s_i$), and commit ($c_i - f_i$) stages of the ROB model
(\Cref{eqn:rob:size,eqn:rob:dep,eqn:rob:mem,eqn:rob:commit}) for $\texttt{ROB} \in \{1, 2, 4, 8, \ldots, 1024\}$.\footnote{
The execution latency does not depend on \texttt{ROB} size; therefore, we only include one copy of the execution latency distribution feature.}
These latency distributions provide additional context that can be useful for understanding certain nuances of the performance dynamics. For example, the execution latency distribution indicates whether a program is load-heavy, which can be useful for predicting memory congestion. 

\subsection{ML Model}
\label{sub:design:ml}
The final component of \ours's design is a lightweight ML model that predicts the CPI of a program on a specified architecture.
The model is a shallow multi-layer perceptron (MLP) (details in \S\ref{sec:implementation}) that takes as input a concatenation of (i) the performance distributions corresponding to the target microarchitecture (\S\ref{subsub:design:per-resource-thr}), (ii) the auxiliary features (\S\ref{sub:design:additional-feat}), and (iii) a 20-dimensional vector of parameters $(\vec{\mathbf{p}})$ representing the target microarchitecture  (\Cref{tab:param-space}). 
We train the ML model on a dataset constructed by randomly sampling diverse program regions and microarchitectures.
We simulate each sample program region and sample microarchitecture using the cycle-level simulator to collect the ground-truth target CPI.
To train the ML model, we use a loss function that measures the relative magnitude of CPI prediction error, as follows:
\begin{equation}
    Loss(\hat{y}, y) = \frac{|\hat{y}-y|}{y},
\label{eqn:loss}
\end{equation}
where $\hat{y}$ denotes the predicted CPI and $y$ denotes the CPI label.
\section{\ours's Implementation Details}
\label{sec:implementation}

\niparagraph{Trace analyzer and analytical models}
We implement the trace analyzer and analytical models in C++.
Trace analysis performs in-order cache simulation (per memory configuration) and branch prediction simulation (for TAGE).
To precompute the performance features for a program, we run
the trace analyzer for each memory configuration to derive the \ours trace, and then run
the analytical models for all parameter values of each CPU resource independently. Our current implementation uses a single thread, but all analytical model invocations could run in parallel.
To calculate performance distributions, \ours uses a window size of $k=400$.

\begin{figure}
    \centering
    \includegraphics[width=\linewidth]{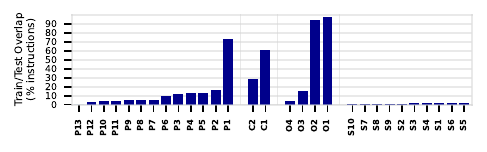}
    \vspace{-6mm}
    \caption{Average test/train overlap across benchmarks}
    \label{fig:overlap}
    \vspace{-3mm}
\end{figure}

\niparagraph{Dataset}
Unless specified otherwise, \ours uses a dataset with $789,024$ data points for training, with an additional $48,472$ unseen (test) data points reserved for evaluation.
Every data point is constructed by independently sampling a microarchitecture, and a $100$k-instruction region.
To sample a microarchitecture, we independently pick a random value from \Cref{tab:param-space} for every parameter.
\ours's large microarchitecture space ($\sim2\times10^{23}$) ensures that test microarchitectures are almost surely unseen during training, preventing memorization.
To sample a program region, we sample a program from \Cref{tab:workload-space},
sample a trace of the chosen program randomly with probability proportional to trace length,
and sample a region randomly from this trace.
\Cref{fig:overlap} shows the average overlap of test program regions with their closest training region (the training region with maximum instruction overlap) for every program.
The overlap is $16.86\%$ on average, and less than $10\%$ for the majority of programs.

\begin{table}[h]
\scriptsize
    \centering
    \caption{
    ML model's 3873-dimensional input.}
    \vspace{-3mm}
    \resizebox{\columnwidth}{!}{
    \begin{tabular}{c|c|c|c}
    \toprule
        \textbf{\makecell{(\S\ref{subsub:design:per-resource-thr}) Per-resource\\throughput analysis}}& \textbf{\makecell{(\S\ref{sub:design:additional-feat})\\Pipeline stalls}} & \textbf{\makecell{(\S\ref{sub:design:additional-feat}) Latency\\distributions}} & \textbf{\makecell{(\Cref{tab:param-space}) Target\\microarchitecture}}\\
    \midrule
      $11\times101=1111$  & $4\times101 + 1 + 11 \times 1 = 416$ & $(1 + 2 \times 11) \times 101 = 2323$ & $19+2\times2 = 23$ \\
    \bottomrule
    \end{tabular}
    }
    \label{tab:input-dim}
    \vspace{-3mm}
\end{table}
\niparagraph{Lightweight ML model}
\label{sub:implementation:ml}
\ours's ML component uses a fully connected MLP with a 3873-dimensional input layer and two hidden layers with sizes $256$ and $128$ that outputs a scalar CPI prediction.
For encoding every input distribution to the ML model, \ours uses a $101$-dimensional encoding which includes $50$ fixed equally-spaced percentiles of the original distribution, $50$ fixed equally-spaced percentiles of the size-weighted distribution,\footnote{The size weighted distribution is a transformation of the original distribution of a non-negative random variable in which we weight every sample by its value. This transform highlights the tail of the original distribution.} and the average value.
\Cref{tab:input-dim} shows the breakdown of the input dimensions to the features detailed in \S\ref{sec:design}.
Note that in the first column, we do not include throughput distributions for static bandwidth resources that remain constant throughout the entire program such as Commit width.
In the last column, we use one-hot vectors for the branch predictor type and the state of prefetching.
We use the AdamW~\cite{adamw} optimizer with weight decay of $0.3$, learning rate of $0.001$ that halves after $\{10, 14, 18, 22\}$k steps, and batch size of $50$k to train for $1521$ epochs.

\section{Evaluation}
\label{sec:eval}

We evaluate \ours's CPI prediction accuracy and speed in \S\ref{sub:eval:acc-speed}. 
In \S\ref{sub:eval:ablation}, we dive deeper into its accuracy and our design choices.

\subsection{\ours's Accuracy and Speed}
\label{sub:eval:acc-speed}
\begin{figure}
    \centering
    \includegraphics[width=\linewidth]{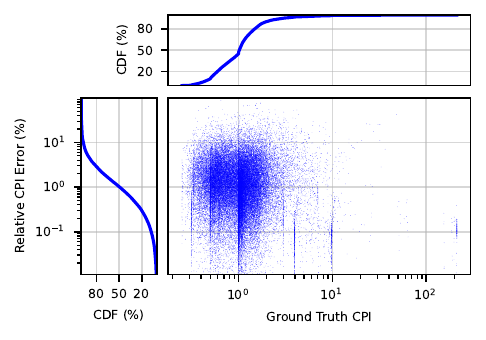}
    \vspace{-5mm}
    \caption{Scatterplot of \ours's CPI prediction error vs. the CPI for unseen (test) pairs of 100k-instruction regions and microarchitectural parameters. The plots on the sides show the distributions of CPI and prediction error.
    The average error is 2\%, with only 2.5\% of samples having larger than 10\% error.
    }
    \vspace{-3mm}
    \label{fig:error-cdf}
\end{figure}

\begin{figure}
    \centering
    \includegraphics[width=\linewidth]{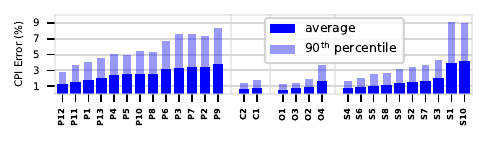}
    \vspace{-6mm}
    \caption{Error breakdown across benchmarks}
    \vspace{-3mm}
    \label{fig:error-breakdown}
\end{figure}

\niparagraph{Accuracy on random microarchitectures}
To highlight the generalization capability of \ours across microarchitectures, we first evaluate its accuracy on the unseen test split of the dataset (\S\ref{sec:implementation}), where microarchitectures are randomly sampled. 
\Cref{fig:error-cdf} illustrates \ours's relative CPI prediction error (\Cref{eqn:loss}) vs. the ground-truth CPI from our gem5-based cycle-level simulator. The top and left plots besides the axes show the distributions of the CPI and \ours's prediction error across all samples.
\ours achieves an average relative error of only $2.03\%$.
Moreover, its error has a small tail; only $2.51\%$ of test samples have errors larger than $10\%$.
Recall from \S\ref{sec:implementation} that such accuracy cannot be achieved by memorization since the microarchitectures in our test dataset are not seen in the training samples.
\Cref{fig:error-breakdown} shows the error breakdown across programs. While some programs are more challenging than others, the average error and P$90$ is capped at $4.2\%$ and $8.9\%$, respectively.
Furthermore, the errors do not correlate well with the per program train/test overlaps in \Cref{fig:overlap}.
For instance, \ours's average error is less than $1\%$ for \workload{S4} and \workload{S6}, and only slightly over $1\%$ for \workload{P12}, all of which have train/test overlaps less than $3.5\%$.
This highlights \ours's effectiveness in generalizing (in distribution) across program regions.

\begin{figure}
    \centering
    \includegraphics[width=\linewidth]{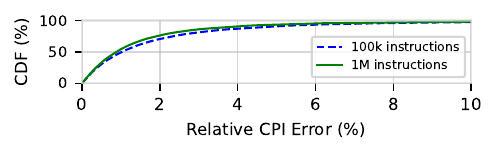}
    \vspace{-7mm}
    \caption{
    \ours is more accurate on longer program regions. 
    }
    \vspace{-3mm}
    \label{fig:1M}
\end{figure}

\niparagraph{Longer program regions}
Recall that one of the main design goals of \ours is to avoid a run time cost that scales with the number of instructions $(\mathcal{O}(L))$.
Hence, unlike cycle-level simulators that operate on sequence of instructions, \ours takes as input a fixed-size performance characterization of the program independent of the program length.
To evaluate \ours on longer program regions, we create a new dataset similar to the original one (\S\ref{sec:implementation}) with longer program regions of $1$M instructions and re-train \ours on it.
\Cref{fig:1M} shows the distribution of \ours's relative CPI prediction error over the unseen test split of this dataset (solid green line).
The average error is $1.75\%$ and only $1.82\%$ of cases have larger than $10\%$ error, which is slightly better than  \ours's accuracy on the original 100k-instruction region dataset (dashed blue line).
We hypothesize that this is because the average CPI has less variability over longer regions due to the phase behaviors getting averaged out (which we confirmed by comparing the CPI variance in the two cases). 
This reduced variance makes the learning task easier for longer regions, boosting \ours's accuracy.

\niparagraph{Accuracy on ARM N1}
To assess \ours's accuracy on a realistic microarchitecture, we evaluate its CPI predictions for ARM N1 (\Cref{tab:param-space}), using the $100$k-instruction regions in the test split of our dataset (\S\ref{sec:implementation}).
It has an average error of $3.25\%$ with $4.39\%$ of program regions having errors larger than $10\%$, which is a slight degradation in the accuracy compared to random microarchitectures.
We believe that this is because randomly sampled microarchitectures are more likely to have a single dominant bottleneck while ARM N1 is designed to be balanced. 

\begin{figure}
    \centering
    \includegraphics[width=\linewidth]{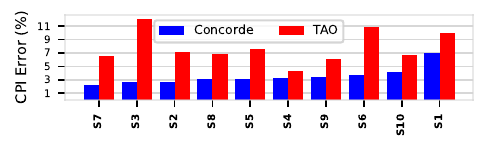}
    \vspace{-7mm}
    \caption{
    \ours is more accurate than TAO on all programs.
    }
    \vspace{-3mm}
    \label{fig:tao-comparison}
\end{figure}

\niparagraph{Comparison with TAO~\cite{tao}}
We compare \ours with TAO, the previous SOTA in sequence-based approximate performance modeling.
Unlike \ours, TAO does not generalize without additional retraining beyond a single microarchitecture.
Hence, we train it for ARM N1 on a dataset of $100$M randomly sampled instructions from SPEC2017 programs (\Cref{tab:workload-space}).
\Cref{fig:tao-comparison} compares TAO's CPI prediction accuracy on $100k$-instruction regions from SPEC2017 programs with \ours's; \ours is more accurate for every single program.
This is despite the fact that \ours is trained on random microarchitectures whereas TAO is specialized to ARM N1.
\begin{figure}
    \centering
    \includegraphics[width=\linewidth]{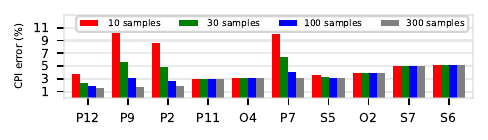}
    \vspace{-7mm}
    \caption{Accuracy for long programs vs. number of samples}
    \vspace{-3mm}
    \label{fig:long-program}
\end{figure}

\niparagraph{Accuracy on long programs}
Using \ours's CPI predictions for finite program regions as the building block, we can estimate the CPI for arbitrarily long programs by randomly sampling program regions and averaging their predicted CPIs. As an example, we use the 1M-instruction region model to predict the CPI for programs with 1B instructions. \Cref{fig:long-program} shows \ours's accuracy in predicting CPI for ARM N1, across ten such 1B-instruction programs, with four sampling levels.
As shown, with as little as $100$ samples, \ours's error gets below $5\%$ for every program, with an average error of 3.5\%. Using 300 samples, the average error decreases to 3.16\%.

\begin{figure}
    \centering
    \includegraphics[width=\linewidth]{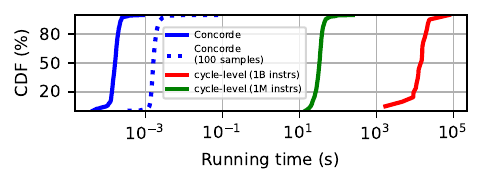}
    \vspace{-7mm}
    \caption{\ours is five/seven orders of magnitude faster than a cycle-level simulator on $1$M/$1$B-instruction program regions.
    }
    \vspace{-3mm}
    \label{fig:time-comparison}
\end{figure}

\niparagraph{\ours's Speed}
\Cref{fig:time-comparison} shows the running time distribution of \ours and our gem5-based cycle-level simulator.
We measure the running time of \ours and the cycle-level simulator on a single CPU core. For these experiments, we simulate from the first instruction of each trace, to avoid extra warmup overheads for the cycle-level simulator. 
The average running time of \ours (solid blue) is $168\mu\sec$.
Compared to our cycle-level simulator, \ours achieves an average speedup of more than $2\times10^5$ for $1$M-instruction regions.
Furthermore, \ours's running time does not change with the length of the instruction region (e.g., $100\text{k}\to1$M) since the size of its input distributions are fixed.
In contrast, the cycle-level simulator's running time scales with the program region length, e.g., $487\times$ by increasing the length from $1$M (green) to $1$B (red) instructions.
Recall that to estimate the CPI of $1$B-instruction programs in \Cref{fig:long-program}, we used \ours's predictions on randomly sampled $1$M-instruction regions. 
The dashed dotted line in \Cref{fig:time-comparison} shows the running time distribution for processing $100$ samples, measured on the same CPU.
Even with 100 sequential samples, \ours's average running time ($1.7m\sec$) is about $10^7$ times faster than the cycle-level simulator for programs with $1$B instructions. 
Additionally, the running time of the cycle-level simulator exhibits a high variance due to its dependence on the number of cycle-level events, which varies with programs and microarchitectures.
In contrast, \ours's running time has minimal variance since its computation is deterministic irrespective of the program or the microarchitecture.
Note that the reported speedups do not include the benefits of batching \ours's calculations on accelerators such as GPUs, which would further amplify its advantage.

\subsection{Deep Dive}
\label{sub:eval:ablation}

\subsubsection{What constitutes \ours's error tail?}
\label{sub:eval:tail-analysis}
Recall from \S\ref{sub:eval:acc-speed} that \ours has a small error tail, where tail is defined as cases with larger than $10\%$ error.
Here, we detail our attempts to understand some of the factors responsible for the tail.

\niparagraph{Discrepancy in raw execution times}
\begin{figure}
    \centering
    \includegraphics[width=\linewidth]{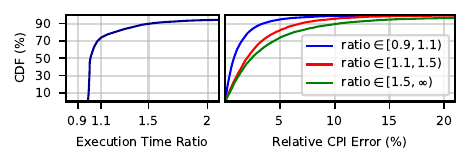}
    \vspace{-7mm}
    \caption{
    Although the ML component of \ours corrects for a large portion of errors in estimates of instruction execution times from trace analysis, this error plays a significant role in the tail of \ours's error distribution.
    }
    \vspace{-3mm}
    \label{fig:discrepancy}
\end{figure}
Recall that \ours's analytical models use approximate instruction execution times derived in trace analysis (\S\ref{sub:design:ta}). 
As we discussed in \S\ref{sub:design:ta}, these estimated executions times can differ from the actual values observed during timing simulations. \Cref{fig:discrepancy} (left) shows the distribution of the ratio of the actual instruction execution times in timing simulations to their estimates from trace analysis, across $100$k-instruction regions in our test dataset (\S\ref{sec:implementation}). 
More than $10\%$ of program regions have a ratio larger than $1.5$. These discrepancies can occur for a variety of reasons, including memory congestion, partial store forwarding, etc. that we do not account for in trace analysis.

With high errors in their raw inputs, our analytical models will be inaccurate. We bucketize program regions based on the above ratio into three buckets, and plot \ours's prediction error distribution for samples in each bucket (\Cref{fig:discrepancy}, right). The result shows that \ours's prediction error increases for the buckets with larger execution time discrepancy. But its accuracy remains quite high, even with significant discrepancies, e.g., achieving an average error of $4.53\%$ in cases with ratio larger than $1.5$. This shows that the ML component of \ours can correct for significant errors in the analytical models. Nonetheless, errors in execution time estimates from trace analysis account for a large portion of the tail of \ours's error distribution. Among test program regions that have errors larger than $10\%$, $41.5\%$ have execution time ratios larger than $1.5$ (whereas only about 10\% of all program regions have a ratio larger than 1.5). 

\niparagraph{Branch prediction}
Recall from \S\ref{sub:design:analytical} that unlike other CPU components, \ours does not analytically model branch mispredictions.
Instead, it relies on a set of auxiliary features that are helpful for learning the effect of pipeline stalls.
We will show in \S\ref{subsub:eval:additional-features} that these features indeed boost \ours's overall accuracy. Here, we study whether branch mispredictions are another source of \ours's error tail.  \Cref{tab:eval:brmiss} categorizes \ours's accuracy based on the number of branch mispredictions in 100k-instruction regions of the test dataset (\S\ref{sec:implementation}).
Intriguingly, \ours's accuracy improves as the number of mispredictions increases, with an average error of \xx{1.82\%} in regions with over 5,000 branch mispredictions.
We hypothesize that this is because programs with large number of stalls have low parallelism and simpler dynamics, making them easier to predict. This result confirms that \ours's branch-related features are sufficient.

\begin{table}
\footnotesize
    \centering
    \caption{\ours successfully learns the effect of branch prediction.}
    \vspace{-3mm}
    \begin{tabular}{c|c c c}
    \toprule
        \textbf{Number of branch mispredictions}& $\mathbf{[0, 1000)}$ & $\mathbf{[1000, 5000)}$ & $\mathbf{[5000, \infty)}$ \\
    \midrule
        \textbf{\ours's average error} $\mathbf{(\%)}$ & $2.16$ & $2.12$ & $\mathbf{1.82}$ \\
        $\mathbf{\%(}$\textbf{\ours's error}$\mathbf{>10\%)}$ & $3.11$ & $2.43$ & $\mathbf{1.95}$ \\
    \bottomrule
    \end{tabular}
    \label{tab:eval:brmiss}
\end{table}

\subsubsection{Ablation study}
\label{subsub:eval:additional-features}
\begin{figure}
    \centering
    \includegraphics[width=\linewidth]{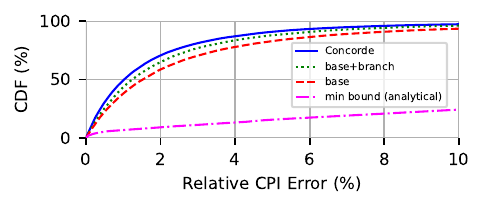}
    \vspace{-7mm}
    \caption{Ablation of \ours's design components
    }
    \vspace{-3mm}
    \label{fig:ablation}
\end{figure}
Recall from \S\ref{sub:design:additional-feat} that \ours uses a few auxiliary features to augment the primary per-component throughput distributions.
We train several variants of \ours to understand the impact of these features. For reference, we begin with a simple minimum over the per-component throughput bounds (no ML). As shown by the pink line in \Cref{fig:ablation}, this has poor accuracy, achieving an average error of $65\%$ (and $11\%$ in cases with no branch misprediction).
\ours's base ML model, which takes as input the per-component throughput distributions along with the branch misprediction rate, significantly boosts accuracy (red line), achieving an average error of $3.32\%$ with errors exceeding $10\%$  in only $4.48\%$ of cases. Further adding the auxiliary features (\S\ref{sub:design:additional-feat}) related to pipeline stalls (green) and instruction latency distributions (blue) provides incremental accuracy improvements, reducing average error to $2.4\%$ and $2.03\%$ and the percentage of samples with errors larger than 10\% to $3.7\%$ and $2.51\%$, respectively.

In addition, we ablated the ML model size, and the choice of $k$, the length of instruction windows for throughput calculations (\S\ref{sub:design:analytical}).
Expanding the model to three hidden layers of sizes 512, 256, and 128 slightly lowers the average error on random microarchitectures from 2.03\% to 1.85\%,
while reducing it to a single hidden layer of size 256 increases the error to 3.91\%.
Varying $k\in\{100, 200, 400\}$ did not have a significant effect on our results.

\subsubsection{Preprocessing cost}
\label{subsub:explore-overhead}
Precomputing the performance features for a 1M-instruction region for all the $2.2\times 10^{23}$ parameter combinations in \Cref{tab:param-space} takes 3959 seconds on a single CPU core\,---\,equivalent to the time required for 107 cycle-level simulations with similar warmup.
This includes 195s for trace analysis (\S\ref{sub:design:ta}) and 3764s for analytical modeling (\S\ref{sub:design:analytical}).
Trace analysis comprises one TAGE, 40 D-cache, and 20 I-cache simulations.
The dominant factors in analytical modeling are $40\times1024$ ROB model invocations (3327s) and $40\times256$ invocations of the Load/Store queue models (211s/211s).
The precomputed performance features occupy 24MB in uncompressed NumPy~\cite{numpy} format.

\Cref{tab:param-space} sweeps all parameters in increments of 1, but such a fine granularity is typically not necessary in practice. Quantizing the parameter space can significantly reduce the precomputation time. For example, considering powers of 2 for ROB, Load and Store queues, i.e, $\texttt{ROB} \in \{1, 2, 4, \ldots, 1024\}$, $\texttt{Load/Store}\text{ queue} \in \{1, 2, 4, \ldots, 256\}$, reduces the analytical modeling time to 63s , lowering the total preprocessing time for the resulting $1.8\times 10^{18}$ parameter combinations to 257s (7 cycle-level simulations). Techniques like QEMU~\cite{qemu} could further reduce trace analysis time~\cite{cache-sim-qemu}.

\subsubsection{Training cost}
\label{subsub:training-overhead}
ML training takes 3 hours on a TPU-v3-8~\cite{tpu-v3} cloud server with 8 TensorCores (2.7 hours on an AMD EYPC Milan processor with 64 cores).
To generate the training dataset (\S\ref{sec:implementation}), we only run trace analysis and analytical modeling for one (randomly selected) microarchitecture for each program region. 

Using 512 cores, it takes 19.4 hours to create the more expensive 1M-instruction region dataset with 837,496 data points. This includes 16.8 hours for cycle-level simulations (to generate the CPI labels), 2.2 hours for trace analysis, and 26 minutes for analytical modeling.

\begin{figure}
    \centering
    \includegraphics[width=\linewidth,
    ]{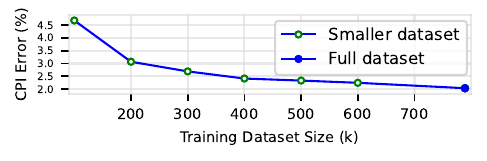}
    \vspace{-7mm}
    \caption{Impact of training dataset size on \ours's accuracy}
    \vspace{-3mm}
    \label{fig:dataset-size-ablation}
\end{figure}

Although training is a one-time cost, it can be reduced at a slight degradation in model accuracy. \Cref{fig:dataset-size-ablation} shows that reducing the training dataset size to 200k samples gradually increases the relative CPI error from 2.01\% to 3.07\%.
Further reduction to 100k samples increases the error to 4.67\%.

\subsubsection{Out-of-Distribution (OOD) Generalization}
\label{subsub:eval:ood}
\begin{figure}
    \centering
    \begin{minipage}{\linewidth}
    \centering
    \begin{subfigure}[t]{\linewidth}
        \centering
        \includegraphics[width=\linewidth]{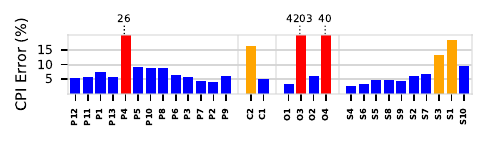}
        \vspace{-4.5mm}
    \end{subfigure}
    \begin{subfigure}[t]{\linewidth}
        \centering
        \vspace{-4.5mm}
        \includegraphics[width=\linewidth]{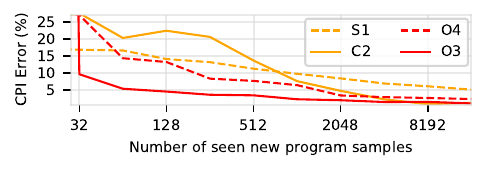}
    \end{subfigure}
    \end{minipage}
    \vspace{-4mm}
    \caption{Errors can be high on unseen programs (\emph{top}). However, \ours recovers quickly as it trains on their samples (\emph{bottom}).}
    \vspace{-3mm}
    \label{fig:ood}
\end{figure}
Like any ML model, we expect \ours to be trained on a diverse dataset of programs representative of programs of interest.
However, to stress test program generalization,
for every program,
we train \ours on a dataset that excludes all its traces and evaluate the accuracy of the resulting model on that program.
\Cref{fig:ood} (top) shows the average OOD error for all programs.
As expected, the error increases, with some programs being affected more than others. 23 programs (blue) have OOD error below 10\%.
The 3 programs with the  highest error (red) are synthetic microbenchmarks testing specific microarchitectural capabilities. These programs are unlike any other in the dataset. For instance, \workload{O3} (a memory test), has much higher CPIs compared to other programs in \Cref{tab:workload-space}.
The 3 remaining programs (orange), with OOD error of about 15\%, are real workloads that stand out from the others.
For example, as we will see in \S\ref{sec:attr}, \workload{S1} has the highest sensitivity to cache sizes among all workloads in \Cref{tab:workload-space}.

Compared to generalization across microarchitectures, OOD generalization across programs is not a major concern. Programs and benchmarks used for CPU architecture exploration are relatively stable. For example, SPEC CPU benchmarks are only updated every few years, and we similarly see infrequent updates to our internal suite of benchmarks. Nevertheless, we quantify the cost of ``onboarding'' new programs into \ours for  \workload{O3}, \workload{O4} (2 highest red bars), and \workload{S1}, \workload{C2} (2 highest orange bars). For each of these programs, we train \ours on all other programs together with a varying number of samples from the new program.
As \Cref{fig:ood} (bottom) shows, 2k (8k) samples from the new program are enough for \ours to reach within 5\% (2\%) of the error floor achieved by the model trained on the full dataset with $\sim30k$ samples per program (\Cref{fig:error-breakdown}).
\workload{O3} and \workload{O4} have the steepest drop in error, which is likely due to the regularity of these synthetic benchmarks.

\subsubsection{Can \ours predict metrics other than CPI?\nopunct}
\label{subsub:eval:other-metrics}
Although we focused on CPI in designing our analytical models, \ours's rich performance distributions are useful for predicting other metrics as well.
To illustrate this point, we retrain \ours's ML model (without changing hyperparameters) to predict the average {\em Rename queue occupancy} (\%) and {\em average ROB occupancy} (\%), on the same dataset used for CPI (\S\ref{sec:implementation}).
On unseen test samples, \ours achieves an average prediction error of 2.50\% and 2.23\%,
respectively, vs. the ground-truth metrics from our gem5-based simulator.
\section{Fine-Grained Performance Attribution}
\label{sec:attr}

Beyond predicting performance, architects often need to understand {\em why} a program performs as it does on a certain design.  In this section, we present a methodology for fine-grained attribution of performance to different microarchitectural components.
Our method can be used in conjunction with any performance model $y = f(\vec{\mathbf{x}}, \vec{\mathbf{p}})$ relating microarchitectural parameters to performance. But as we will see, it is computationally impractical for expensive models such as cycle-level simulators. \ours's massive speedup over conventional methods makes such large-scale, fine-grained analyses possible.

Concretely, our goal is to quantify the relative impact of different microarchitectural parameters $\vec{\mathbf{p}}$ on the performance of a program $\vec{\mathbf{x}}$. This requires identifying the dominant performance bottlenecks. Many existing performance analysis techniques (e.g., Top-Down~\cite{top-down}, CPI stacks~\cite{cpi-stack}) rely on hardware performance counters to identify bottlenecks. We seek to obtain similar insights using only a performance model $y = f(\vec{\mathbf{x}}, \vec{\mathbf{p}})$ like \ours that outputs the (predicted) performance of a program given the microarchitectural parameters. 
Performance of a single microarchitecture $\vec{\mathbf{p}}$ provides no information about which of the parameters $p_i$ are important. Thus, we use parameter {\em ablations}, where we change some parameters and observe their impact on performance. Intuitively, parameters that have a large effect on performance when modified are  more important. Parameter ablations are commonly used to understand the impact of design choices~\cite{vidushi1, vidushi2, vidushi3, vidushi4}. A typical approach is to start with microarchitectural parameters $\mathbf{p}^{base}$ representing a {\em baseline} design, and modify one parameter (dimension) at a time to reach a {\em target} design with parameters $\mathbf{p}^{target}$. After each parameter change, the incremental change in performance is reported as the contribution of that parameter to the total performance difference between $\mathbf{p}^{base}$ and $\mathbf{p}^{target}$.

\begin{figure}
    \centering
    \includegraphics[width=\linewidth]{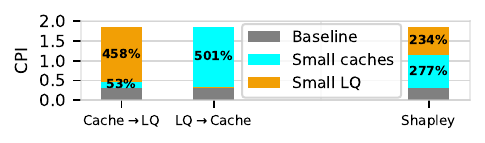}
    \vspace{-7mm}
    \caption{Changing the order of parameter ablations (Cache $\rightarrow$ Load Queue vs. Load Queue $\rightarrow$ Cache) leads to different conclusions about their relative importance. Shapley values provide a fair, order-independent performance attribution to design parameters.
    }
    \vspace{-3mm}
    \label{fig:strawman}
\end{figure}

Although this methodology is standard, it can be difficult to draw sound conclusions from parameter ablations when there are multiple inter-related factors effecting performance. The issue is that the {\em order} of parameter ablations can change the perceived importance of different factors. To illustrate, suppose we are interested in quantifying the relative impact of (i) limited cache size and (ii) limited Load queue size on a memory-intensive workload. As a baseline, we consider a ``big core'' with all parameters set to their largest value in \Cref{tab:param-space}, particularly: L1d/L1i cache = 256kB, L2d cache = 4MB, and Load queue = 256. \Cref{fig:strawman} shows the CPI achieved by this baseline (Grey bars) on a sample trace from the \workload{Search3} workload.

Next, we consider two parameter ablations atop the baseline, where we reduce the cache sizes and the Load queue size to our target values: L1d/L1i cache = 64kB, L2d cache = 1MB, Load queue =~12. In one ablation, we first reduce the cache sizes and then the Load queue size; in the other ablation, we reduce the Load queue size first, then reduce cache size. 
The two left bars in \Cref{fig:strawman} show the CPI trajectory for both routes, along with the CPI increase associated with reducing cache sizes and Load queue size in each case. The overlaid numbers show the percentage CPI increase relative to the baseline following each parameter change. 

The two orders of parameter ablations lead to entirely different conclusions. The (Cache $\rightarrow$ Load queue) order suggests that reducing the Load queue size has about $9\times$ larger impact than reducing the cache sizes. The (Load queue $\rightarrow$ Cache) order, on the other hand, says that reducing the Load queue has negligible effect and the performance degradation is almost entirely caused by reducing the cache sizes. Neither of these interpretations is correct. The reality is that the effects of cache and Load queue size are intertwined. A large Load queue can mitigate the performance hit of small caches for this workload (due to increased parallelism). Similarly, a large cache size can perform well despite a small Load queue (since Load instructions complete quickly). It is only when {\em both} the Load queue and cache sizes are small that we incur a large performance hit.

\niparagraph{Shapley value: a fair, order-independent attribution} A natural way to remove the bias caused by a specific order of parameter ablations is to consider the average of {\em all} possible orders. Let $\Pi$ denote the set of all permutations of the parameter indices $D\triangleq\{1,\ldots,d\}$. Each permutation $\pi \in \Pi$ corresponds to one order of ablating the parameters from $\mathbf{p}^{base}$ to $\mathbf{p}^{target}$, resulting in a different value for the incremental effect of modifying parameter $i$. 

Specifically, define $\mathbf{p}_\pi(j) \triangleq (\mathbf{p}_{\pi_{1:j}}^{target}, \mathbf{p}_{\pi_{j+1:d}}^{base})$ to be the $j^{th}$ microarchitecture encountered in the ablation study based on order $\pi$, i.e., parameters $\pi_1, \ldots, \pi_j$ are set to their target values and the rest remain at the baseline. Let $k$ denote the position of parameter $i$ in the order $\pi$. Then, the incremental effect of parameter $i$ in order $\pi$ is: $\Delta_i^\pi \triangleq f(\mathbf{x}, \mathbf{p}_\pi(\text{k})) - f(\mathbf{x}, \mathbf{p}_\pi(\text{k-1}))$. To assign an overall attribution to parameter $i$, we take the average over all permutations:
\begin{equation}
\varphi_i \triangleq \frac{1}{|\Pi|} \sum_{\pi \in \Pi} \Delta_i^\pi,
\label{eqn:shapley}
\end{equation}
where $|\Pi| = d!$ is the total number of permutations.

The quantity defined in \Cref{eqn:shapley} is referred to as the {\em Shapley value}~\cite{shapley} in economics. The concept arises in cooperative game theory, where a group of $M$ players work together to generate value $v(M)$. Shapley's seminal work showed that the Shapley value is a ``fair'' distribution of $v(M)$ among the players, in that it is the only way to divide $v(M)$ that satisfies certain desirable properties (refer to~\cite{shapley} for details). In our context, the ``players'' are the different microarchitectural components, and the ``value'' to be divided is the performance difference between the baseline and target microarchitectures.\footnote{It is not difficult to see that: $\sum_i \varphi_i = f(\mathbf{x}, \mathbf{p}^{target}) - f(\mathbf{x}, \mathbf{p}^{base}).$} The Shapley value is used in many areas of science and engineering~\cite{Shapley1,Shapley2,Shapley3,Shapley4,Shapley5,Shapley6,Shapley7}, but to our knowledge, we are the first to apply it to performance attribution in computer architecture.

The rightmost bar in \Cref{fig:strawman} shows the Shapley values corresponding to cache and Load queue sizes in the above example. The Shapley value correctly captures that small caches and small Load queue sizes are together the culprit for high CPI relative to the baseline, with a slightly larger attribution to small caches.

\niparagraph{Case study}
To illustrate Shapley value analysis, we use it for fine-grained performance attribution in a target design based on the ARM N1 core~\cite{arm-n1} (parameters in \Cref{tab:param-space}) across our entire pool of programs. As baseline, we use the ``big core'' configuration mentioned above (perfect branch prediction, other parameters set to their max).

Computing Shapley values is computationally expensive. For each program (region), using Eq.~\eqref{eqn:shapley} directly requires $d! \times d$ performance evaluations, where $d$ is the number of parameters. We can calculate an accurate Monte Carlo estimate of Eq.~\eqref{eqn:shapley} using a few hundred randomly sampled permutations, but even that requires a massive number of performance evaluations for large-scale analyses. For example, estimating Shapley values for our corpus of workloads (\Cref{tab:workload-space}) using 2000 sample regions per program and 200 permutations of parameter orders requires $\sim$143M CPI evaluations in total. This is impractical with existing cycle-level simulators; we estimate it would take about a month on a 1024-core server! With \ours, the computation takes about an hour on a TPU-v3~\cite{tpu-v3} cloud server with 8 TensorCores.

\begin{figure}
    \centering
    \includegraphics[width=\linewidth]{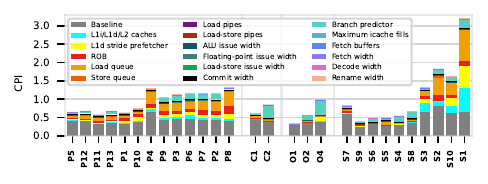}
    \vspace{-7mm}
    \caption{CPI attribution for ARM N1 across all workloads
    }
    \vspace{-3mm}
    \label{fig:attr}
\end{figure}

\Cref{fig:attr} shows the result of our analysis.
The grey bars show the reference CPI achieved by the ``big core'' baseline, while the entire bars show the CPI achieved by ARM N1.
Within each workload group, i.e., proprietary, cloud, open-source, and SPEC2017, the programs are sorted based on the relative CPI increase of ARM N1 compared to the baseline. 
For instance, in SPEC2017 benchmarks, \workload{S1(505.mcf\_r)} has the largest relative jump in CPI for ARM N1, whereas  \workload{S7(557.xz\_r)} has the smallest relative CPI increase.

The colored bars in \Cref{fig:attr} show the Shapley value for each microarchitectural component, i.e., how much each component in ARM N1 is responsible for the performance degradation relative to the baseline. This provides a bird's eye view of the dominant performance bottlenecks across the entire  corpus of workloads. For instance, all the proprietary programs and half of the SPEC2017 programs are mainly backend bound on ARM N1, with prominent bottlenecks being the Load queue size and ROB size.
A few programs such as \workload{S4(541.leela\_r)} (a chess engine using tree search) are frontend bound, with the TAGE branch predictor the most prominent frontend bottleneck. Cache sizes and L1 prefetching have a large effect on some SPEC2017 benchmarks (e.g., \workload{S10(502.gcc\_r)}, \workload{S1}) but a less pronounced impact on our proprietary workloads, perhaps in indication of our programs being cache-optimized.

\begin{figure}
    \centering
    \includegraphics[width=\linewidth]{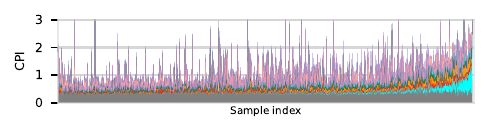}
    \vspace{-6mm}
    \caption{CPI attributions for all \workload{Search3(P9)} sample regions
    }
    \vspace{-3mm}
    \label{fig:zoom}
\end{figure}

For a deeper look, we can further zoom into the behavior of a single program. 
For example, \Cref{fig:zoom} shows the CPI attribution for all $2000$ sample regions of \workload{P9}, sorted on the x-axis based on their sensitivity to cache size.
Although the \workload{P9} bar in \Cref{fig:attr} shows limited sensitivity to cache sizes on average, the zoomed-in view in \Cref{fig:zoom} shows high sensitivity to cache size in about $10\%$ of the sampled regions, highlighting the different phase behaviors in the program~\cite{rev-D}.
\section{Related Work}
\label{sec:related}
\niparagraph{Conventional CPU simulators}
Conventional simulators~\cite{ye2000design,hughes2002rsim,austin2002simplescalar,ortego2004sesc,ubal2007multi2sim,yourst2007ptlsim,chiou2007fpga,miller2010graphite,genbrugge2010interval,patel2011marss,rico2011trace,carlson2011sniper,sanchez2013zsim,ahn2013mcsima+,alves2015sinuca,elrabaa2017very,akram2019survey,brais2020survey,gober2022championship} aim to balance speed and accuracy, leveraging higher abstraction levels~\cite{genbrugge2010interval,carlson2011sniper}, or decoupled simulations of core and shared resources~\cite{sanchez2013zsim,elrabaa2017very}.
While these methods achieve faster simulations, they often compromise flexibility or accuracy. 
Hardware-accelerated simulators~\cite{karandikar2018firesim,chiou2007fpga} improve speed but require extensive development effort for validation.
Statistical modeling tools~\cite{simpoint,wunderlich2003smarts,sherwood2001basic,sherwood2003phase,eeckhout2005exploiting,joshi2006measuring,guo2015accelerating} reduce computation by sampling representative segments~\cite{simpoint,wunderlich2003smarts} or generating synthetic traces~\cite{eeckhout2003statistical,nussbaum2001modeling}, but they trade off flexibility and detailed insights.

\niparagraph{Analytical performance models}
Analytical models~\cite{analytical,abel2023facile,van2015micro,chen2011hybrid,analytical-reviewerC-1,analytical-reviewerC-2,mindgap} provide quick performance estimates using parameterized equations and microarchitecture-independent profiling. 
Such methods are ideal for crude design space exploration but often lack the granularity needed to capture intricate $\mu$-architectural dynamics. 

\niparagraph{ML- and DL-based performance models}
Conventional ML based models~\cite{joseph2006construction,lee2006accurate,lee2007illustrative,ipek2006efficiently,li2009machine,zheng2016accurate,joseph2006predictive,dubach2007microarchitectural,seshadri2022evaluation,wu2022survey} predict performance over constrained design spaces but often struggle with fine-grained program-hardware interactions. 
In contrast, \ours demonstrates robust generalization across unseen programs and microarchitectures.
Recent DL-based models~\cite{mendis2019ithemal,sykora2022granite,perfvec,tao,pandey2022scalable,chaudhary2024comet,apollo} improve modeling at a higher abstraction levels at the cost of higher compute.
Notably, PerfVec\cite{perfvec} (significant training overhead) and TAO\cite{tao} (additional finetuning for unseen configurations) emphasize per-instruction embeddings. 
Our work diverges by compactly capturing program-level performance characteristics using analytical models and fusing them with a lightweight ML model for capturing dynamic behaviors.
\section{Final Remarks}
\label{sec:conclusion}
\label{sec:discussion}

The key lesson from \ours is that decomposing performance models into simple analytical representations of individual microarchitectural components, fused together by an ML model capturing higher-order complexities, is very effective. It enables a method that is both extremely fast and accurate. Before concluding, we remark on some limitations of our work and directions for future research. 

\ours does not obviate the need for detailed simulation. It enables large-scale design-space explorations not possible with current methods (e.g., Shapley value analysis (\S\ref{sec:attr})), but some analyses will inevitably require more detailed models. Moreover, \ours needs training data to learn the impact of design changes (e.g., different parameters), which we currently obtain using a reference cycle-level simulator. 
In principle, \ours could be trained on data from any reference platform, including emulators and real hardware. 

As an ML approach, \ours's accuracy is inherently statistical. Our results show high accuracy for a vast majority of predictions, but there is a small tail of cases with high errors. We have analyzed some of the causes of these errors (\S\ref{sub:eval:tail-analysis}), and we believe that further improvements to the analytical models (e.g., explicitly modeling in-memory congestion) can further reduce the tail. But we do not expect that tail cases can be eliminated entirely. Alternatively, a large set of techniques exist for quantifying the uncertainty of such ML models~\cite{conformal1,conformal2,uncertainty}. Future work on providing confidence bounds would allow designers to detect predictions with high potential errors and crosscheck them with other tools.

Finally, \ours was just one example of our compositional analytical-ML modeling approach. We believe that the methodology is broadly applicable and we hope that future work will extend it to other use cases, such as modeling multi-threaded systems, uncore components, and other architectures (e.g., accelerators). 
\begin{acks}
We thank Steve Gribble and Moshe Mishali for their comments on earlier drafts of the paper.
We thank Jichuan Chang, Brad Karp, and Amin Vahdat for discussions and their feedback.
We thank Derek Bruening, Kurt Fellows, Scott Gargash, Udai Muhammed, and Lei Wang for their help in running cycle-level simulations.
We also thank the extended team at SystemsResearch@Google and Google DeepMind who enabled and supported this research direction.
\end{acks}
\vfill
\bibliographystyle{isca2025}
\bibliography{refs}

\end{document}